 \documentclass[preprint2]{aastex}
\usepackage{ctable}
\usepackage{amssymb}

\shorttitle{Fossil signatures in the thick disk}
\shortauthors{Curir et al.}

\begin{document}

\title{The radial metallicity gradients  in the Milky Way thick disk \\
    as fossil signatures of a primordial chemical distribution.}

\author{A. Curir}
\affil{INAF - Osservatorio Astrofisico di Torino - Italy}
\email{curir@oato.inaf.it}
\author{A.L. Serra}
\affil{ INAF - Osservatorio Astrofisico di Torino - Italy\\
Universit\`a di Torino, Dipartimento di Fisica, Torino - Italy\\
Istituto Nazionale di Fisica Nucleare (INFN), Sezione di Torino, Torino - Italy}
\author{A. Spagna}
\affil{INAF - Osservatorio Astrofisico di Torino - Italy}
\author{M.G.  Lattanzi}
\affil{INAF - Osservatorio Astrofisico di Torino - Italy}
\author{P. Re Fiorentin}
\affil{INAF - Osservatorio Astrofisico di Torino - Italy}
\author{A. Diaferio}
\affil{Universit\`a di Torino, Dipartimento di Fisica, Torino - Italy\\
Istituto Nazionale di Fisica Nucleare (INFN), Sezione di Torino, Torino - Italy}

\begin{abstract}
In this letter we examine the evolution of the radial metallicity gradient induced by secular processes, in the disk of an $N$-body Milky Way-like galaxy. We assign a [Fe/H] value to each particle of the simulation according to an initial, cosmologically motivated, radial chemical distribution  and let the disk dynamically evolve for $\sim 6$ Gyr. This direct approach allows us to take into account only the effects of dynamical evolution and to gauge how and to what extent they  affect the initial chemical conditions.  

The initial [Fe/H] distribution increases with $R$ in the inner disk up to $R \approx 10$ kpc and decreases for larger $R$. We find that the initial chemical profile does not undergo major transformations after $\sim 6$ Gyr of dynamical evolution. The final radial chemical gradients predicted by the model in the solar neighborhood are positive and of the same order of those recently  observed  in the Milky Way thick disk. 

We conclude that:
\item(1) the spatial chemical imprint at the time of disk formation is not washed out by secular dynamical processes, and 
\item(2) the observed radial gradient may be the dynamical relic of a thick disk originated from a stellar population showing a positive chemical radial gradient in the inner regions.
\end{abstract}


\section{Introduction}

 Observations suggest that the negative radial metallicity gradient, $d$[Fe/H]$/dR$, measured in the solar neighborhood close to the plane, flattens or even attains positive values
if we select stars at higher $|z|$ from the galactic plane or with high [$\alpha$/Fe] abundances (Allende Prieto et al. 2006; Boeche et al. 2013, Anders et al. 2013; Hayden et al. 2013). 
This effect can be a signature of a positive radial gradient in the thick disk population, as derived by Marsakov \& Borkova (2005) and by Carrell et al. (2012), who analyzed kinematically selected thick disk samples.
Alternatively, a similar feature can be simply produced by the variation of the fraction of thin disk and thick disk stars due to their different scale-height and scale-length (Anders et al. 2013). This scenario is indeed consistent with the short scale-length, $h_R\sim 2$ kpc, recently estimated for $\alpha$-enhanced stars by Bensby et al. (2011) and Cheng et al. (2012). 

Furthermore, there is observational evidence indicating that the Milky Way (MW) is a barred galaxy. Non-axisymmetric structures in the disk (such as bars and spiral arms) drive galaxy evolution: they redistribute angular momentum among the different galactic components and can enhance radial migration inside the disk. It is still matter of debate the amount of impact that  these secular  processes can have  on the formation of the thick disk and on its chemical and kinematical properties \citep{roskar2008, loebman2011, brunetti11, minchev12, Kubryk13}.  

This work is aimed at analyzing whether dynamical effects can be responsible for the non-negative radial metallicity gradients observed today in the MW thick disk,  and suggests that such gradient can be regarded as  a cosmological feature.

We use an $N$-body disk that develops a bar and whose final population displays a radial metallicity gradient; 
this is obtained from an initial axisymmetric disk with a positive radial metallicity gradient for $R<10$ kpc, which has dynamically evolved for 6.1 Gyr. This secular evolution model was recently proven \citep{Curir2012} to reproduce the observed velocity-metallicity correlation in the thick disk \citep{spagna2010, lee2011}. In this letter, we show that the positive slope of the chemical radial gradient in the MW thick disk observed today in the solar neighborhood might be a signature remaining from the primordial population of disk stars. 

\section{Numerical simulations and chemical tagging}

The simulation used in this work was presented in  \cite{Curir2012};
it is an exponential disk with structural parameters appropriate for a MW-like DM halo at redshift $z=0$.
The disk,  having a mass of $5.6{\cdot}10^{10} M_{\odot}$ and scale length h=3.5 kpc,  is embedded in a Navarro, Frenk and White (NFW) dark matter halo \citep{NFW97} with $10^7$ particles  having mass $M_{vir} = 10^{12} M_{\odot}$  and  concentration $C_{vir}=7.4$  \citep[see][for details]{Curir2012}. 
 We ran the simulation using the public parallel treecode GADGET2 \citep{sprin2005} on the cluster matrix at the CASPUR (Consorzio Interuniversitario per le Applicazioni del Supercalcolo) consortium, Rome.
Our simulation can be considered as an evolving thick disk model, consistent with the scenario where the thick disk is originated from gas accreted in a chaotic period at high redshift and with short star formation timescales \citep{Brook04, Brook12}. The initial radial metallicity distribution of the disk can be seen as the link between our isolated simulated disk and the cosmological scenario, as we explain in the following paragraphs.

\begin{figure*}
\centering
\includegraphics[angle=0,scale=0.8,trim=0 0 0 0]{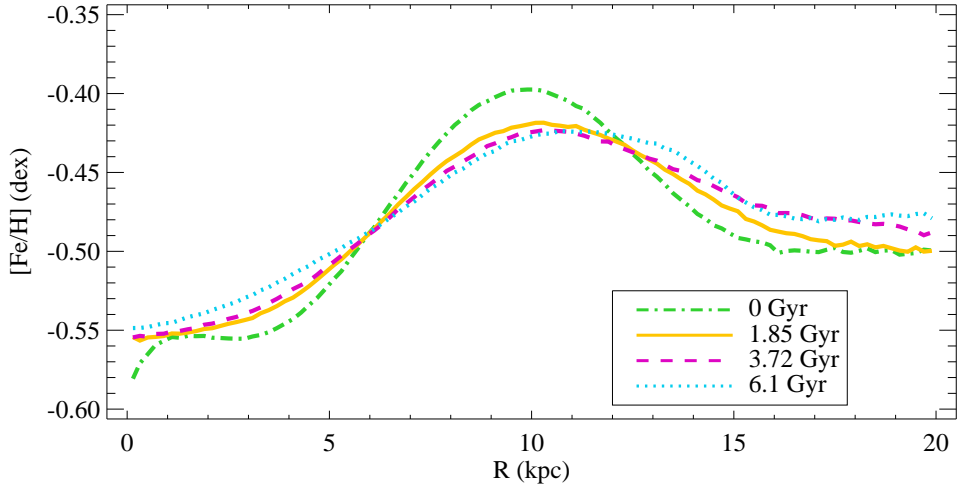}
\caption{Median radial chemical distribution at the beginning of the dynamical simulations (dotted-dashed line), at $\approx$ 1.9 Gyr (solid line), 3.7 Gyr (dashed line) and 6.1 Gyr (dotted line). The points are the median values of the data over bins 0.2 kpc  wide.}
\label{fig:initialdist}
\end{figure*}

In order to avoid the effect of relaxation of the initial conditions, we let the dynamical system evolve for 1 Gyr. Then the time is set to zero and each particle in the disk is tagged with a [Fe/H] label, according to the initial radial function shown in Figure \ref{fig:initialdist}. This function is based on  Model S2IT from \cite{spit01}. We assign metallicities with a dispersion $\sigma_{\rm [Fe/H]}=0.1$ dex, which represents the variation of the median chemical profile over 1 Gyr, according to \cite{spit01}.  Regarding the metallicity distribution as a function of the height, $z$, from the galactic plane, no initial vertical gradient has been included in our simulation.

The positive slope of the function for $R<10$ kpc is justified with an inside-out disk formation model: at early epochs the efficiency of chemical enrichment in the inner regions of the disk is low  due to a large amount of infalling primordial gas. This scenario is consistent with recent results published by \cite{cresci2010}, who found evidence for an ``inverse'' (i.e. positive) metallicity gradient in a sample of galaxies with stellar mass $>3\times 10^9 \, M_{\odot}$, at $z\sim3$. They conclude that this is produced by the accretion of primordial gas \citep{Mott13}, which diluted the abundance of elements heavier than helium in the center of the galaxies.

\begin{figure*}
\includegraphics[angle=0,scale=0.8,trim=0 0 0 0]{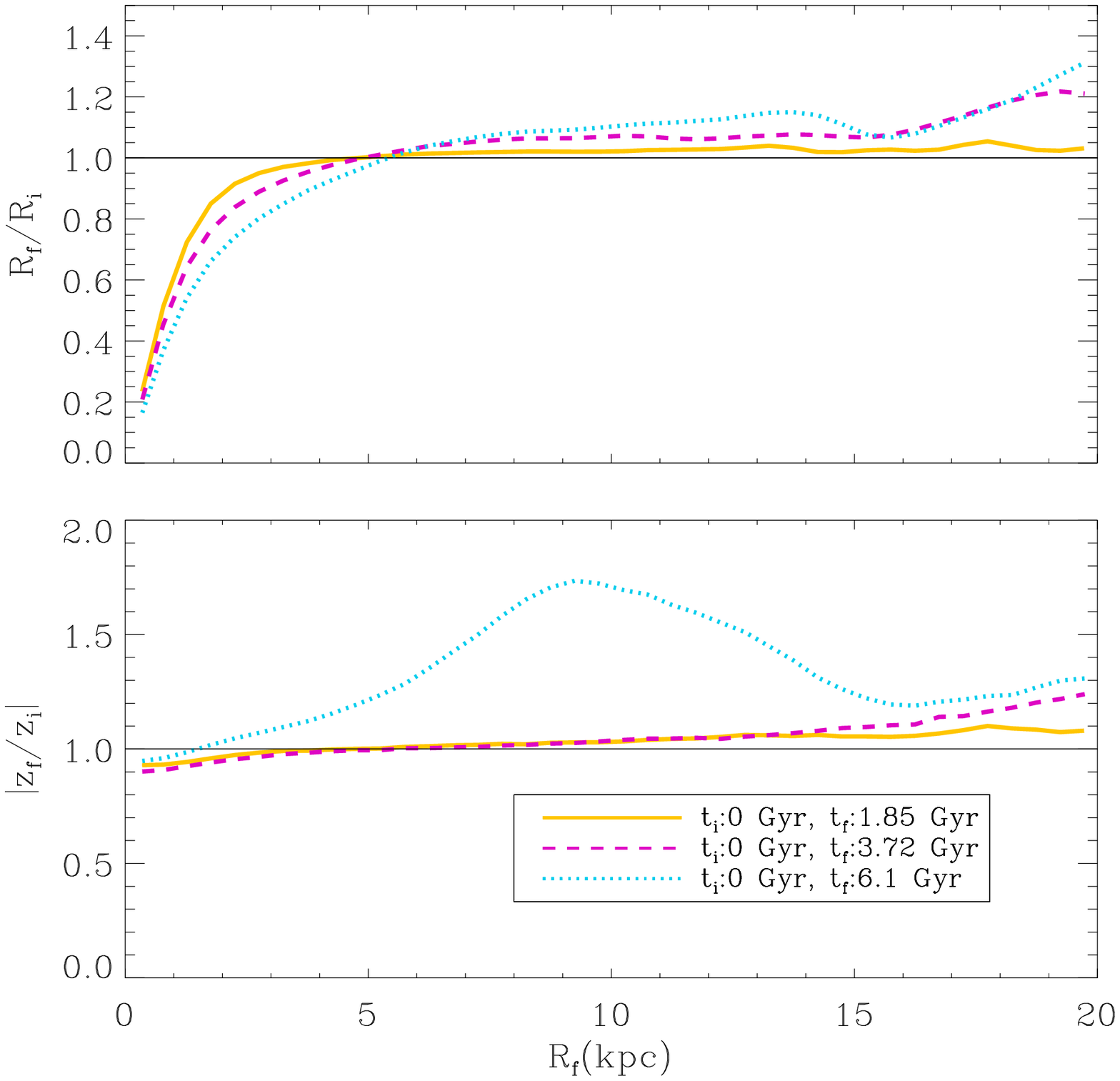}
\caption{Migration of particles. Top panel: profiles of the median ratio $R_{\rm f}/R_{\rm i}$, where $R_{\rm i}$ is the radius of a particle at initial time and $R_{\rm f}$ is that of the same particle at 1.85 Gyr (solid line), 3.7 Gyr (dashed line) or 6.1 Gyr (dotted line). Bottom panel: profiles of the median ratio $|z_{\rm f}/z_{\rm i}|$, at the same times. The binning range over the x-axis is 0.5 kpc in both panels.}
\label{fig:migration}
\end{figure*}

The simulation produces a bar at around 2 Gyr which persists throughout. The radial migration, illustrated in Figure \ref{fig:migration}, redistributes the angular momenta of the orbits inside and outside the bar. In Figure \ref{fig:migration} we can see to what extent the radius of each particle in the simulation has changed from $R_i$, at the start of the dynamical time, to $R_f$: 1.9, 3.7 or 6.1 Gyr later. The stars with $R_{f}/R_{i}<1$ have moved towards inner regions, whereas the other stars have migrated further away. Thus, while the particles in the inner parts have a tendency to go towards the center, the outer regions of the disk are expanding. As for the distribution perpendicular to the plane of the disk, we observe a similar behaviour: the internal regions tend to  compact with time along $|z|$, whereas, in the external regions the disk particles increase their median $|z_{f}/z_{i}|$ \citep[e.g.][]{loebman2011, roskar2013}. The profile $|z_{f}/z_{i}|$ at 6 Gyr is the result of  the formation of a boxy bulge, probably due to buckling instabilities, although further analysis is needed in order to provide a full consistent explanation.

%

\section{Results and discussion}

Figure \ref{fig:initialdist}  shows the median radial chemical distribution at four different dynamical epochs: $t=0$, 1.9,  3.7  and 6.1 Gyr. The radial migration flattens out the initial chemical profile in the disk outer region, while it does not induce major modifications in the inner regions. The peak in the metallicity distribution, initially at $9.9$ kpc, shifts to $10.7$ kpc after $\sim$ 6 Gyr; also the value of the  [Fe/H] peak varies from an initial value of $-0.40$ dex kpc$^{-1}$ to $-0.42$ dex kpc$^{-1}$. These relatively small changes point out the fact that dynamics alone can only marginally affect the initial radial chemical distribution.

\begin{figure*}
\includegraphics[angle=0,scale=0.8,trim=0 0 0 0]{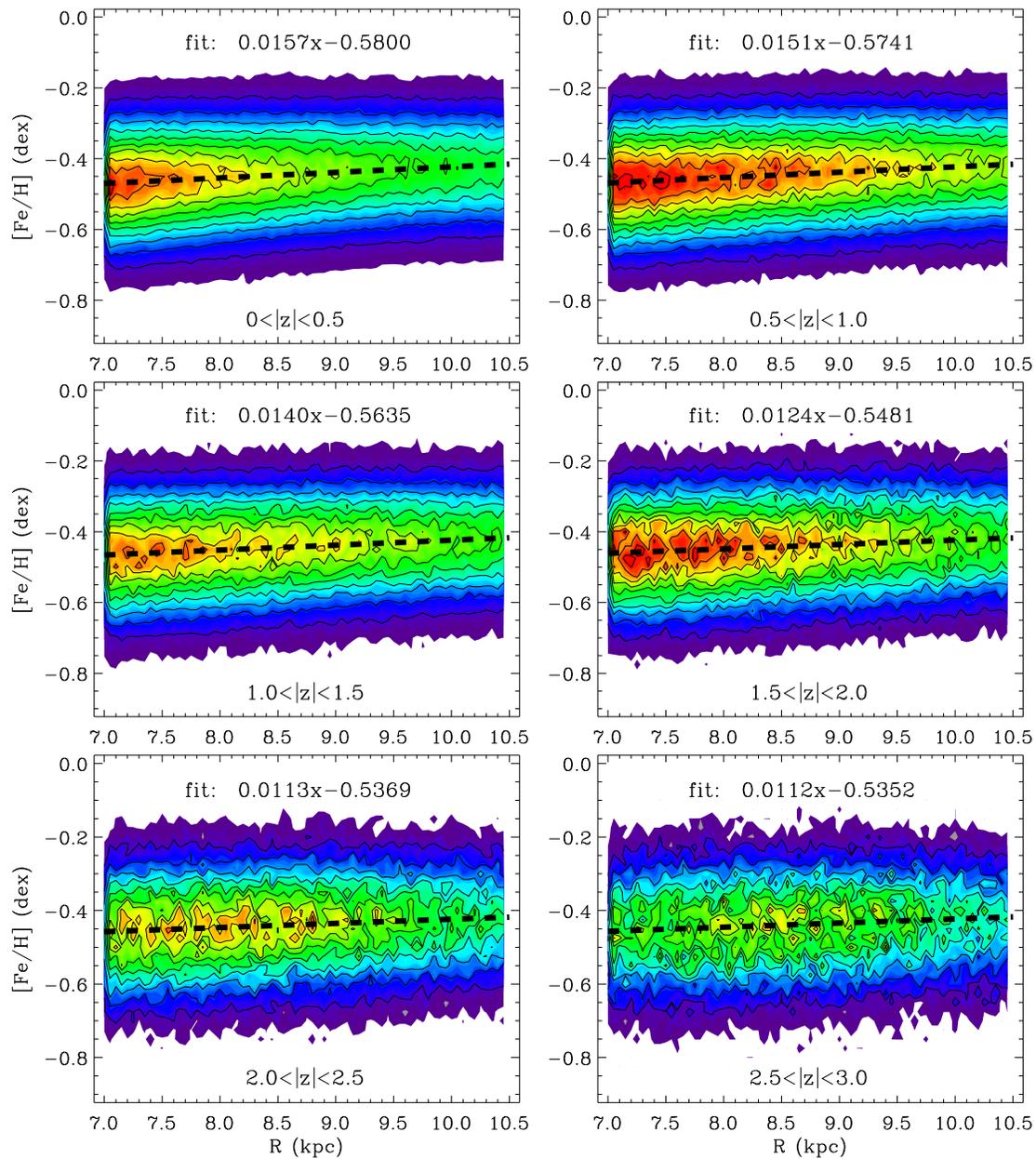}
\caption{Radial chemical distribution of the disk stars in a few kpcs of the MW solar circle, i.e. 7 kpc$<R< $ 10.5 kpc, after 6.1 Gyr of dynamical evolution. The different panels refers to different $|z|$ ranges. The black dashed lines are the best fit to the median values of [Fe/H] in each bin in R (of 0.05 kpc).}
\label{fig:metdist}
\end{figure*}

\begin{figure*}
\includegraphics[angle=0,scale=0.8,trim=0 0 0 0]{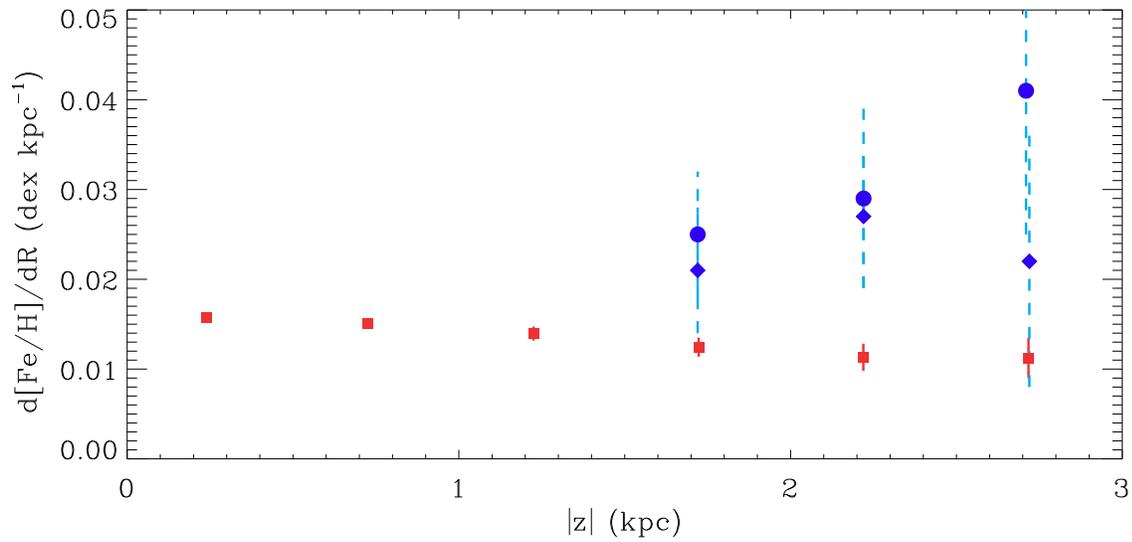}
\caption{Radial metallicity gradients of the disk stars with $7$ kpc $< R < 10.5$ kpc. The  squares are our estimates of the metallicity gradients at 6.1 Gyr from the dynamical $N$-body simulation. The abscissa of each point are the mean height  of the particles in the respective bin (broad $0.5$ kpc in $R$ or $|z|$). The error bars on the vertical axis are 3-$\sigma$ width of the distribution, as 1-$\sigma$ errors would be hardly appreciated in this figure. Circles and diamonds are the gradients  for $|z| > 1.5$ kpc measured by \cite{Carrell12} through the Isochrone Distance and the Photometric Distance, respectively.}
\label{fig:gradients}
\end{figure*}

Figure \ref{fig:metdist} shows the radial chemical distribution of the simulated disk particles  after  6.1 Gyr of dynamical evolution. The particles are taken within a few kpc from the solar circle  (between $7$ kpc and $10.5$ kpc), at different intervals of $|z|$. We find a positive slope that slightly decreases  with $|z|$ from $d{\rm [Fe/H]}/dR = 0.0157 \pm 0.0002$ dex kpc$^{-1}$ for $|z| <$ 0.5 kpc to $0.0112\pm 0.0007$ dex kpc$^{-1}$  in the range 2.5 kpc $< |z| < $ 3.0 kpc (the squares in Figure \ref{fig:gradients}). We have measured $d{\rm [Fe/H]}/dR$ by calculating the median of the chemical distribution in bins of $0.05$ kpc in $R$ and fitting a straight line to those medians, taking into account the 1-$\sigma$ width of the distribution of the data in each bin; the gradient uncertainties are taken as the 1-$\sigma$ error of the fit to the medians.

Recently, \cite{Carrell12} examined the metallicity distribution (in $R$ and $|z|$) in the Galactic thick disk, using F, G and K dwarf stars selected from SDSS DR8.   We compare our simulation with their sample of kinematically selected thick disk stars. Here we consider only the results derived by Carrell et al. (2012) for $|z|>1.5$ kpc, in order to minimize the contamination of thin disk stars (Bensby et al. 2011).

\cite{Carrell12} measured the metallicity gradient in the radial directions for stars in the annular region  ($7.0$ kpc $< R < 10.5$ kpc from the galactic center) and at several distances from the galactic plane.  For thick disk stars they obtained an increasing $d{\rm [Fe/H]}/dR$ with $|z|$, from $0.025\pm0.007$ dex kpc$^{-1}$ for $1.5$ kpc$ <|z|<2 $ kpc to $0.041\pm0.016$ dex kpc$^{-1}$ for  $2.5$ kpc $<|z|<3.0$ kpc (Isochrone Distance Method). These results  are consistent, within 1-$\sigma$ uncertainties, with a constant positive slope of $\approx 0.032$ dex kpc$^{-1}$ for the Isochrone Distance Method and a slope of  $\approx 0.023$ dex kpc$^{-1}$ for the Photometric Distance Method. 

 A mild positive gradient of $0.010 \pm 0.005$  dex kpc$^{-1}$  was also measured  by \cite{Hayden13}, among $\alpha$ -enhanced stars with $T_{\rm eff}> 4200  K $ located between  $|z|=1$ kpc and  $|z|=2$ kpc.  Our results suggest a mildly decreasing positive slope of the order of $\sim 0.013$ dex kpc$^{-1}$.

The evolution of the radial gradients in the solar neighborood is illustrated in Figure \ref{fig:grad_evol}. The initial radial gradients are given by the model we use to assign the metallicities and, as there is no initial dependence on $|z|$, the radial gradients at different $|z|$ have approximately the same values. These profiles reach a substantial stability between $t\sim 1.8$  and 6 Gyr, for 0 kpc $ <|z|< $ 2.5 kpc.

 After 6 Gyr of evolution of our model galaxy, we do not see any obvious trend in the vertical metallicity gradients as a function of radial distance from the Galactic center. We find a significant, but tiny, positive $d{\rm [Fe/H]}/d|z| $ of $0.0031\pm 0.0003$ dex kpc$^{-1}$ for the stars in the radial and vertical ranges $7.0$ kpc $< R < 10.5$ kpc and $1$ kpc $< |z| <3$ kpc .
On the other hand \cite{Carrell12} measured $d{\rm [Fe/H]}/d|z| = -0.113\pm0.010$ dex kpc$^{-1}$ for the whole sample of thick disk stars in  the same spatial region.  \cite{Hayden13} measured a value of $-0.22 \pm 0.02$  dex kpc$^{-1}$ for  $d{\rm [M/H]}/d|z| $ in the range  $7$ kpc $< R <9$ kpc and of $-0.13 \pm 0.04$  dex kpc$^{-1}$  in the range  $9$ kpc $< R <11$ kpc.


\section{Conclusions}

The crucial role of a positive  slope $d{\rm [Fe/H]}/dR$ in the inner disk for the primordial chemical distribution to produce a positive rotation-metallicity correlation in the thick disk \citep{spagna2010, lee2011} was already shown in \cite{Curir2012}.
In this work we have analyzed how the same cosmologically motivated chemical distribution allows us to constrain the influence of the dynamical evolution on the metallicity gradients observed today in the MW thick disk.

We have drawn the radial and vertical chemical distributions from the same simulated MW galaxy used in \cite{Curir2012}, which includes a {\it positive} radial chemical gradient in the inner disk.  At 6.1 Gyr, within the solar neighborhood ($7.0$ kpc $ < R < 10.5$ kpc),  we still obtain a positive slope for the radial chemical distribution in the inner region. This slope  decreases  in going from $|z| < 0.5$ kpc to the region $2.5$ kpc $< |z| <3$ kpc ($d{\rm [Fe/H]}/dR = 0.016\pm0.002$ dex kpc$^{-1}$ and $d{\rm [Fe/H]}/dR = 0.011\pm0.007$ dex kpc$^{-1}$, respectively). The radial chemical gradients predicted by the model are at most within a factor of 2 of those observed by \cite{Carrell12}, with the Photometric Distance Method, and  by \cite{Hayden13}. Our  final positive radial gradients are  a direct consequence of our initial conditions, that include a positive radial gradient in the inner disk.
 
\begin{figure*}
\includegraphics[angle=0,scale=0.8,trim=0 0 0 0]{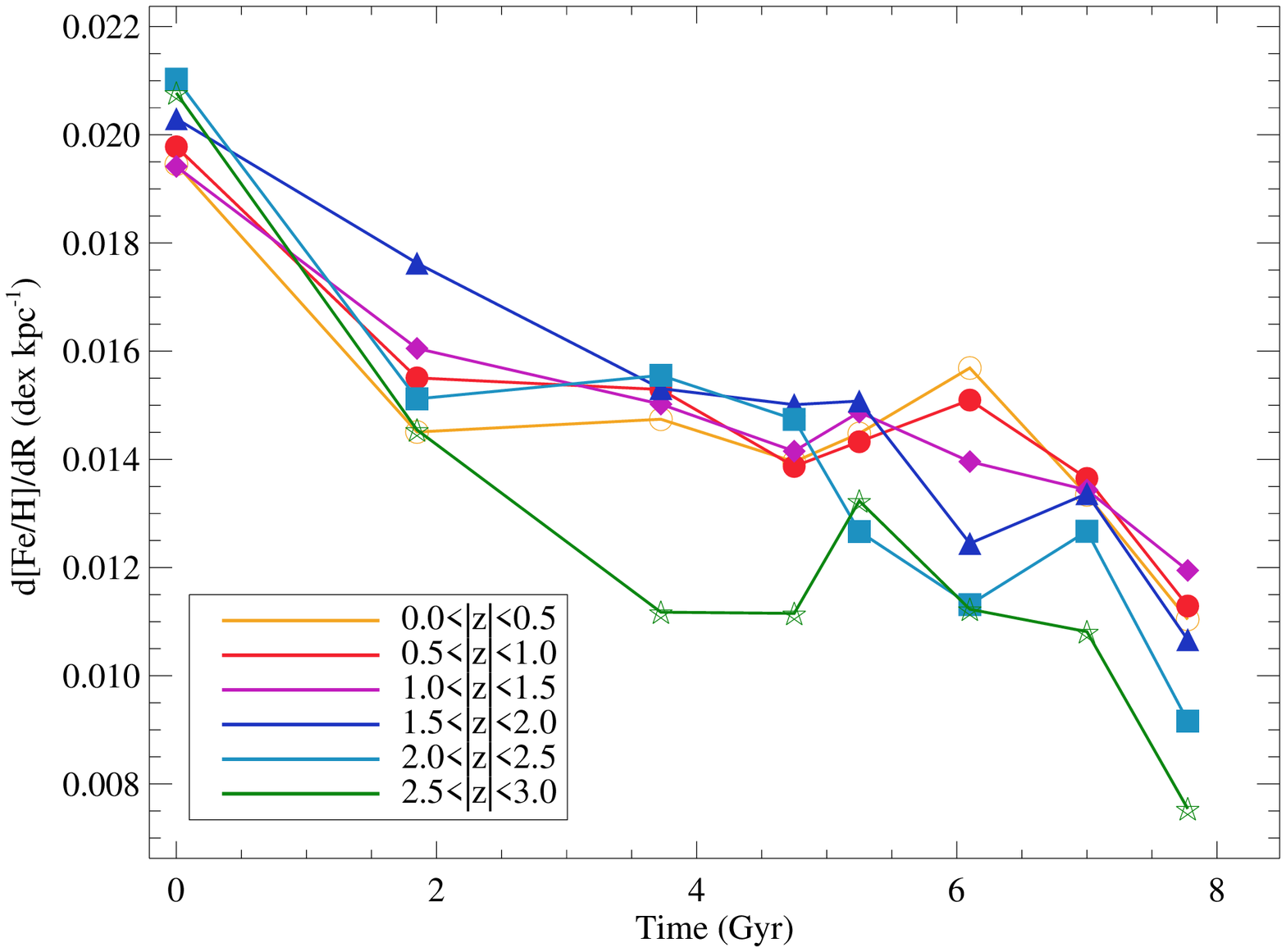}
\caption{Evolution of the radial gradient in the solar annulus in several layers in $|z|$, namely $0$ kpc $<|z|< 0.5$ kpc (empty circles), $0.5$ kpc $<|z|< 1.0$ kpc (filled circles), $1.0$ kpc $<|z|< 1.5$ kpc (diamonds), $1.5$ kpc $<|z|< 2.0$ kpc (triangles), $2.0$ kpc $<|z|< 2.5$ kpc (squares), $2.5$ kpc $<|z|< 3.0$ kpc (stars).}
\label{fig:grad_evol}
\end{figure*}

As shown in Figure \ref{fig:grad_evol}, the radial gradients  in the solar neighborhood are of the same order from 1.8 Gyr to 6.1 Gyr, independently of the layer in $|z|$ (although some redistribution appears to take place).  Such a stability seen between $2$ and $6$ Gyr is consistent with the stability of the rotation-metallicity correlation pointed out by \cite{Curir2012} in the same time range.

Furthermore, the vertical gradient in  the final configuration of our simulation is  very close to zero. This originates from the fact that we do not impose any dependence on $|z|$ in the initial chemical distribution. The lack of a $|z|$-dependence is also responsible from the lack of a trend for the radial gradients to increase with $|z|$.

  We conclude that  the secular disk evolution does {\it not} appear to be able to modify significantly the disk global chemical profiles, suggesting that the radial and vertical chemical distribution of the thick disk is likely to be a fossil signature of the original distribution.
 
Our results indicate that,   if the positive radial metallicity gradient in the solar neighborhood will be confirmed by the observations of  the  Milky Way  thick disk, this is  consistent with a thick disk population showing an  early positive [Fe/H] radial gradient in the inner disk ($R < 10$ kpc) and negative in the outer disk ($R > 10$ kpc). In the context of the inside-out formation scenarios of the galactic disk \citep{spit01,Mott13}, such a gradient ``inversion'' derives from the strong infall of primordial gas that can occur at early times in the inner disk.

\acknowledgments 

The authors thank R. Drimmel for accurate insightful comments.

We are grateful to F. Matteucci and E. Spitoni for providing us with the data of their chemical gradient model for a primordial disk.
AC, PRF, AS acknowledge the financial support of the PRIN-MIUR 2012, grant No 1.05.01.97.02, ``Chemical and dynamical evolution of our Galaxy and of the galaxies of the Local Group''.

ALS and AD acknowledge partial support from the INFN grant InDark and from the grant Progetti di Ateneo/CSP TO Call2 2012 0011 of the Universit\`a di Torino.

MGL and ALS aknowledge partial support from MIUR and the Italian Space Agency through contract no. I/058/10/0 (The Italian participation in  the Gaia Mission).

\end{document}